\documentclass[
 apl,
 aip,
preprint,
]{revtex4-2}

\usepackage{multirow}
\usepackage{float}
\usepackage{graphicx}
\usepackage{dcolumn}
\usepackage{bm}
\usepackage{multirow}
\usepackage[mathlines]{lineno}

\usepackage[utf8]{inputenc}
\usepackage[T1]{fontenc}
\usepackage{mathptmx}
\usepackage{etoolbox}
\usepackage[english]{babel}

\makeatletter
\def\@email#1#2{%
 \endgroup
 \patchcmd{\titleblock@produce}
  {\frontmatter@RRAPformat}
  {\frontmatter@RRAPformat{\produce@RRAP{*#1\href{mailto:#2}{#2}}}\frontmatter@RRAPformat}
  {}{}
}
\makeatother

\begin{document}

\title{Radio frequency single electron transmission spectroscopy of a semiconductor Si/SiGe quantum dot}

\author{I. Fattal}
  \email{imri.fattal@kuleuven.be}
\author{J. Van Damme}

\affiliation{ 
  IMEC, Kapeldreef 75, 3001 Leuven, Belgium
}
\affiliation{ 
  Department of Electrical Engineering, KU Leuven, 3001 Leuven, Belgium
}
\author{B. Raes}
\author{C. Godfrin}
\author{G. Jaliel}
\author{K. Chen}
\affiliation{ 
  IMEC, Kapeldreef 75, 3001 Leuven, Belgium
}
\author{T. Van Caekenberghe}
\author{A. Loenders}
\affiliation{ 
  IMEC, Kapeldreef 75, 3001 Leuven, Belgium
}
\affiliation{ 
  Department of Electrical Engineering, KU Leuven, 3001 Leuven, Belgium
}%
\author{S. Kubicek}
\author{S. Massar}
\author{Y. Canvel}
\author{J. Jussot}
\author{Y. Shimura}
\affiliation{
  IMEC, Kapeldreef 75, 3001 Leuven, Belgium
}
\author{R. Loo}
\affiliation{
  IMEC, Kapeldreef 75, 3001 Leuven, Belgium
}
\affiliation{
Department of Solid-State Sciences, Ghent University, Krijgslaan 281, building S1, 9000 Ghent, Belgium
}
\author{D. Wan}
\author{M. Mongillo}
\affiliation{
  IMEC, Kapeldreef 75, 3001 Leuven, Belgium
}
\author{K. De Greve}
\affiliation{ 
  IMEC, Kapeldreef 75, 3001 Leuven, Belgium
}%
\affiliation{ 
  Department of Electrical Engineering, KU Leuven, 3001 Leuven, Belgium
}%
\affiliation{ 
  Proximus Chair in Quantum Science and Technology, KU Leuven, 3001 Leuven, Belgium
}%

\date{\today}

\begin{abstract}
  Rapid single shot spin readout is a key ingredient for fault tolerant quantum computing with spin qubits.
  An RF-SET (radio-frequency single electron transistor) is predominantly used as its the readout timescale is far shorter than the spin decoherence time. 
  In this work, we experimentally demonstrate a transmission-based RF-SET using a multi-module semiconductor-superconductor assembly. 
  A monolithically integrated SET placed next to a double quantum dot in a Si/SiGe heterostructure is wire-bonded to a superconducting niobium inductor forming the impedance-transforming network. 
  Compared to RF reflectometry, the proposed set-up is experimentally simpler without the need for directional couplers. 
  Read-out performance is benchmarked by the signal-to-noise (SNR) of a dot-reservoir transition (DRT) and an interdot charge transition (ICT) in the double quantum dot near the SET as a function of RF power and integration time. 
  The minimum integration time for unitary SNR is found to be 100 ns for ICT and 300 ns for DRT. 
  The obtained minimum integration times are comparable to the state of the art in conventional RF reflectometry set-ups. 
  Furthermore, we study the turn-on properties of the RF-SET to investigate capacitive shifts and RF losses. 
  Understanding these effects are crucial for further optimisations of the impedance transforming network as well as the device design to assist RF read-out. 
  This new RF read-out scheme also shows promise for multiplexing spin-qubit readout and further studies on rapid charge dynamics in quantum dots.
\end{abstract}

\maketitle

\section{\label{sec:level1}Introduction}
Conventional Single Electron Transistor (SET) electrometers offer significant potential for applications in quantum sensing and quantum computing because of their high sensitivity \cite{morello_single-shot_2010,pla_single-atom_2012}. 
However, in practical use, the bandwidth of conventional SETs is restricted to a few kilohertz. This limitation arises from the high capacitance of the connecting cables (typically of the order of $C_{cable}\sim 1nF$) that link the device's output to room temperature electronics. 
Additionally, at these low frequencies, the charge sensitivity is degraded by relatively large 1/f noise caused by the motion of background charges \cite{elsayed_low_2024,shehata_modeling_2023}.\\

In 1998, Schoelkopf et al. introduced the radio-frequency SET (RF-SET), capable of measuring the charge state of quantum dots with a bandwidth exceeding 100 MHz \cite{schoelkopf_radio-frequency_1998}. 
Their approach involved placing the SET at the end of a low-impedance RF transmission line (50 Ohm), while recording the reflected phase and amplitude --- now dependant on the SET's impedance --- of an applied RF signal.
The key innovation for achieving high-bandwidth measurements consisted of inserting an impedance transformer between the high-impedance SET and the standard $Z_0 = 50\Omega$ characteristic impedance of the measurement lines, thereby addressing the impedance mismatch problem. 
Since then, RF techniques for quantum dots have advanced significantly \cite{vigneau_probing_2023}, with widespread applications in quantum sensing and quantum computing. 
Their high bandwidth has enabled groundbreaking studies of fast charge dynamics in quantum dots and demonstrated high-fidelity readout of silicon-based spin qubits on short timescales \cite{connors_charge-noise_2022,connors_rapid_2020,takeda_fault-tolerant_2016}.\\

In this paper, we investigate the impedance of an SET, monolithically integrated into a Si/SiGe platform, by measuring the transmission through a capacitively coupled transmission line. 
Unlike conventional reflectometry, only a few studies have employed a transmission-type of setup. 
In Ref. \cite{fujisawa_charge_2000}, the SET is embedded in an LC resonator with a resonance frequency of 630MHz and the transmitted signal through the resonator is measured. 
They demonstrated the basic functionality in an AlGaAs/GaAs heterostructure reaching a charge sensitivity of $5\times10^{-4} e/\sqrt{Hz}$ at their maximum modulation bandwidth of 10kHz, as limited by the bandwidth of the lines. 
The SET in that experiment has been used to study a nearby charge trap, however, no signal-to-noise study of charge transitions on a capacitively coupled quantum dots has been performed. 
Moreover, the implementation doesn’t allow for frequency multiplexing. 
In Ref. \cite{zheng_rapid_2019,hamonic_combining_2024} a transmission type setup is employed for multiplexed gate-based reflectometry readout which probes the charge susceptibility, distinguishing whether or not an electron can oscillate between the dots in response to the probe power. 
Despite reaching a signal-to-noise ratio (SNR) of about six within an integration time of only 1 µs in Ref. \cite{zheng_rapid_2019}, gate-based reflectometry readout is more complex to tune up than the conventional RF-SET reflectometry method. \\

The transmission-based setup introduced in Section \ref{sec:level2} of this work measures the transmitted signal past a capacitively coupled load, where the load impedance is the combined impedance of the single electron transistor and an impedance transforming network. 
It offers several advantages over the conventional reflection-based approach: (1) In reflection mode, the reflected signals can interfere with the original signal, introducing noise, loss, and distortion on the SET receiver side. 
(2) Transmission mode does not require additional components, such as a directional coupler, to separate the incident and reflected signals. 
The presence of these separating components leads to unavoidable signal loss in reflection. 
(3) The transmission resonance appears as a dip on a flat baseline, which is beneficial for multiplexed readout. 
Additionally, the transmitted signal at resonance is a monotonic function of the load impedance and therefore less sensitive to exact parameter matching. 
For all of these reasons, we have implemented a superconductor-semiconductor multi-module microwave assembly to demonstrate radio frequency single electron transmission spectroscopy readout. \\

In Section II, we present the fundamental principles of RF SET transmission spectroscopy. 
Section III examines how various circuit parameters influence the performance of the RF single electron transistor transmission spectroscopy setup, and we compare the performance to a conventional reflection mode measurement. 
Section IV.I addresses the effects of microwave losses on the performance of RF-SET readout. 
Finally, in Section IV.II, we implement an RF-SET transmission spectroscopy measurement scheme to investigate a Si/SiGe quantum dot device, demonstrating a projected minimum integration time of $t_{min}<1\mu s$ to achieve a signal-to-noise ratio SNR=1 for electron transitions, which is comparable to the state of the art in conventional reflectometry-based approaches\cite{volk_fast_2019}\cite{liu_radio-frequency_2021}

\section{\label{sec:level2}Radio frequency single electron transistor transmission spectroscopy}

\begin{figure}[H]
  \centering
  \setlength\tabcolsep{1.5pt}
  \begin{tabular}{c c}
    (a) SEM image
    &
    (b) RF Circuit
    \\
    \includegraphics[scale=0.05]{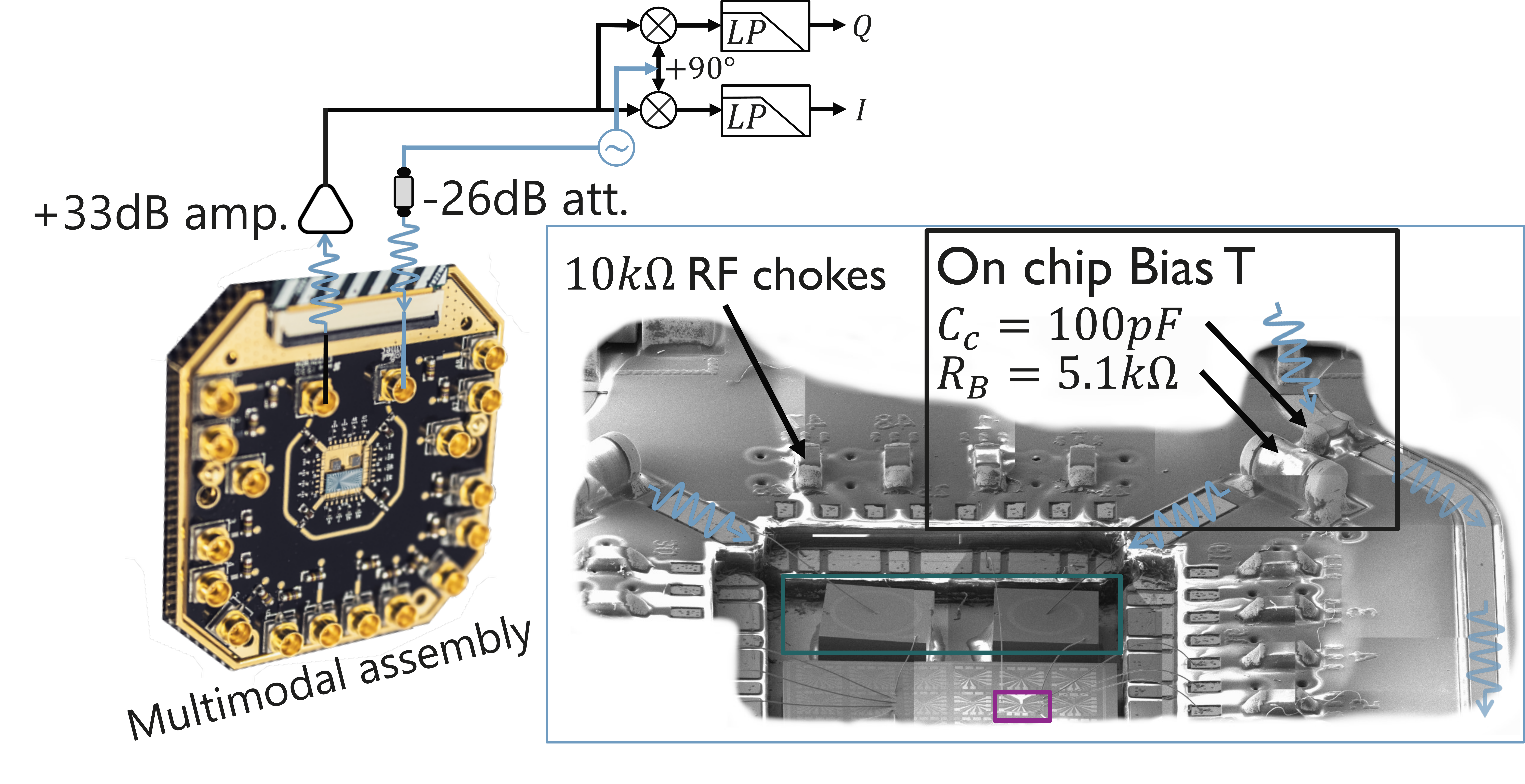}
    &
    \hspace*{-0.5cm}\includegraphics[scale=0.05]{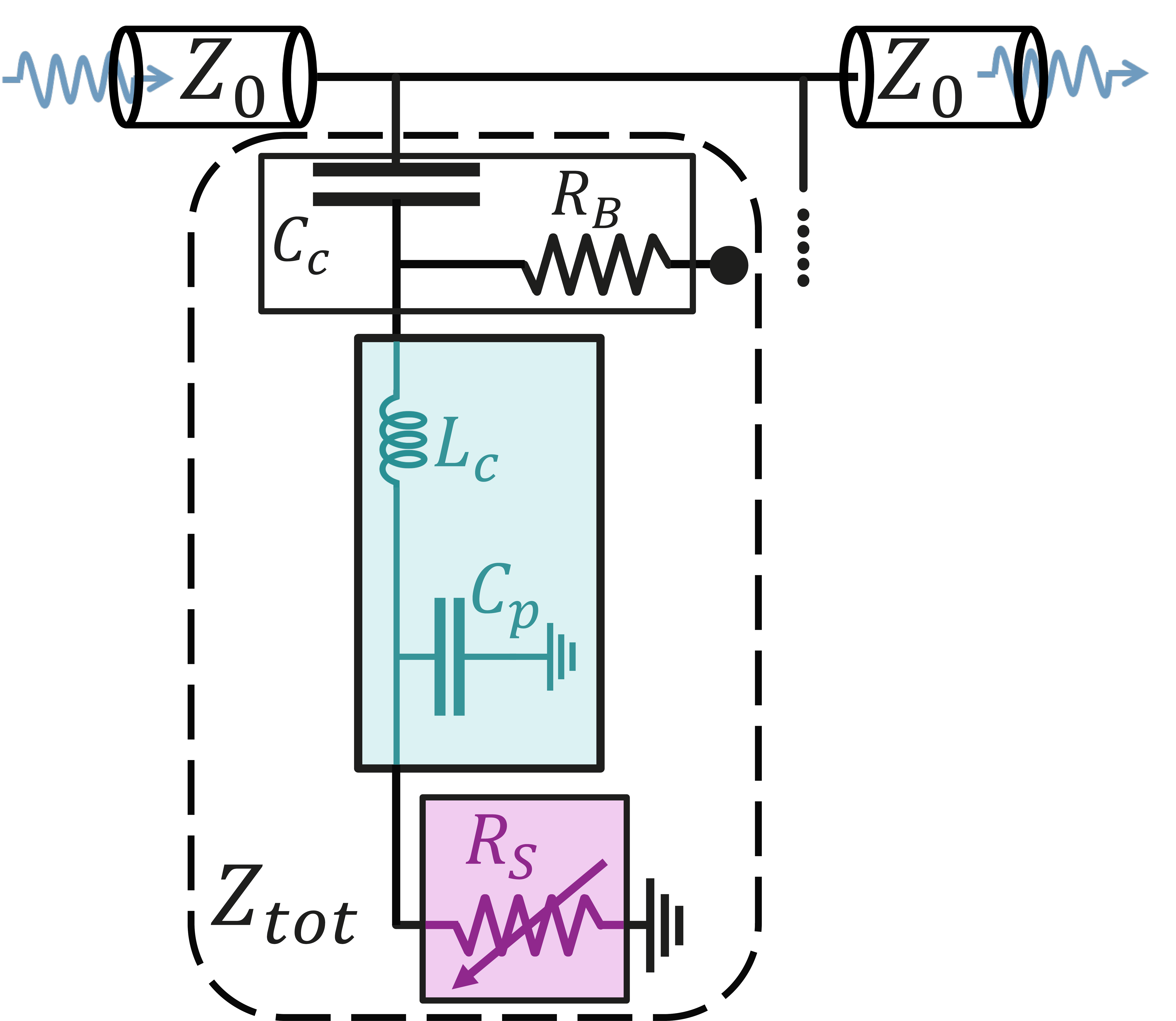}
    \\
  \end{tabular}
  \\
  \setlength\tabcolsep{1.5pt}
  \begin{tabular}{c c}
    (c) Impedance transforming circuit
    &
    (d) SET on sample
    \\
    \includegraphics[scale=0.06]{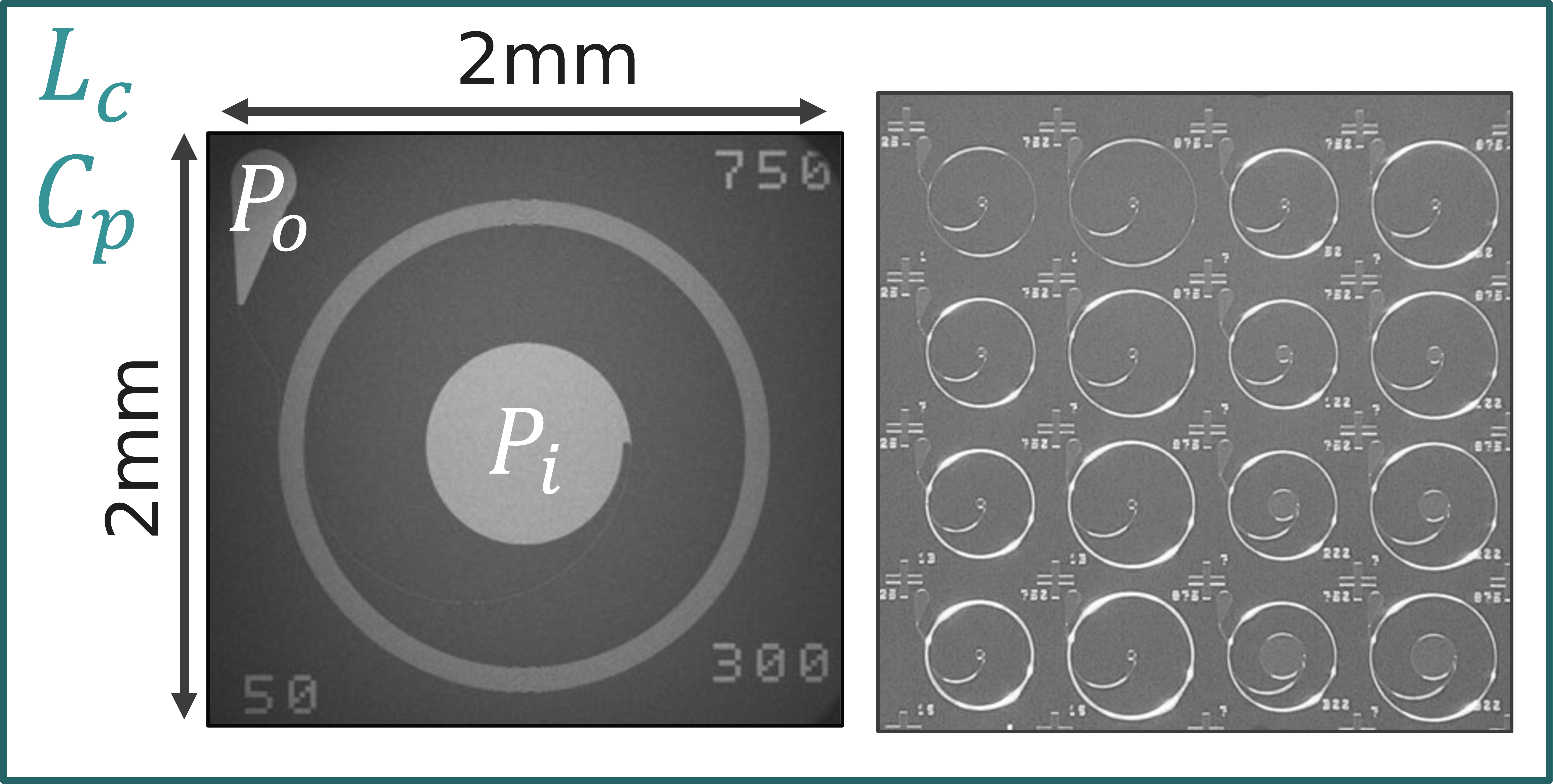}
    &
    \includegraphics[scale=0.05]{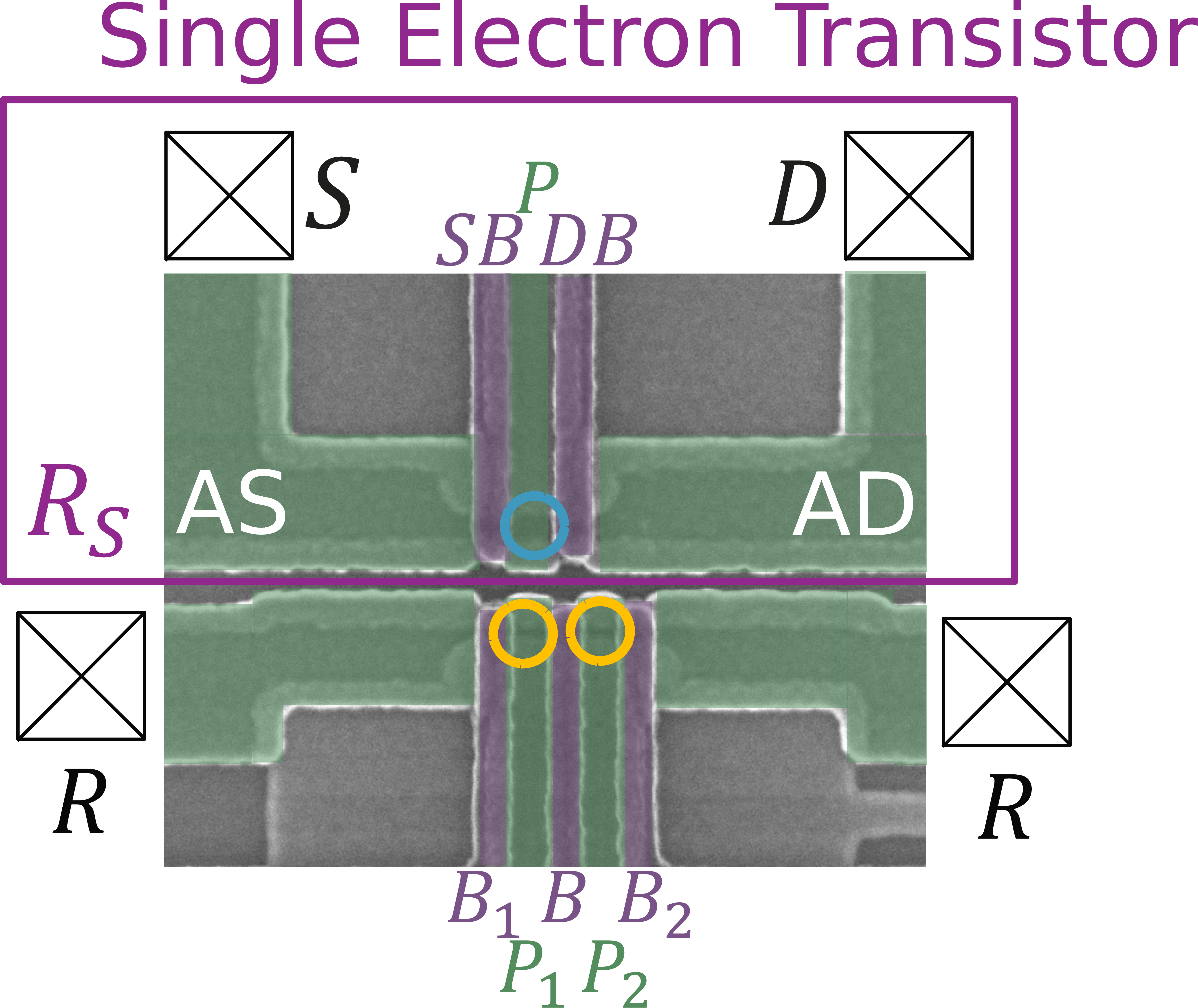}
    \\
    
  \end{tabular}
  \caption{(a) The circuit is implemented using a multimodal approach. 
  A coplanar waveguide is located on the printed circuit board (PCB). 
  Four surface mount coupling capacitors $C_c=100pF$ allow to couple four impedance transforming circuits for multiplexed readout. 
  A $5.1k\Omega$ surface mount resistor allows to dc bias the Ohmic implant of the SET. 
  The impedance circuit consists of a superconducting planar spiral inductor. 
  A wide range of niobium inductors are fabricated using a 300mm optical industrial lithography process and a subtractive etching process. 
  The inductances range from $L_C: 0.004-23\mu H$ and are diced in 2x2mm dies. 
  Conventional wire-bonding is used to: (i) connect the centre pad of the spiral inductor ($P_i$) to the contact pad connected to the coupling capacitor and (ii) connect the b) outer pad of the spiral inductor ($P_o$) to the Ohmic source contact of a single electron transistor (SET) located on a separate die. 
  The transmission response in terms of I and Q values is measured using a conventional demodulation procedure, as schematically indicated. 
  (b) A lumped-element schematic of the circuit for RF-SET transmission spectroscopy. 
  (c) Scanning electron microscopy (SEM) images of niobium inductors (1 on the left and a collection of different designs on the right) forming the impedance transforming circuit ($L_C$  and $C_P$). 
  The inductor has an outer pad $P_o$, used to connect to the source contact of the SET and an inner pad $P_i$ connected to the contact pad on the PCB. 
  (d) A false coloured SEM image of a Si/SiGe device consisting of a single electron transistor (indicated by the blue circle) capacitively coupled to a double quantum dot (indicated by the two yellow dots) and a SEM image of the spiral inductors.}
  \label{fig:1}
\end{figure}

The circuit for RF-SET transmission spectroscopy is depicted schematically in Fig. \ref{fig:1}(b). 
In this configuration, we consider a single SET capacitively coupled to a feedline through a coupling capacitor $C_C$ and connected via an impedance transforming circuit. 
This transmission-based setup can be easily extended to accommodate multiple SETs, facilitating multiplexed readout. 
The phase and amplitude of the transmitted signal $S_{21}$ depend on the ratio of the impedances $Z_0/Z_{tot}$, where $Z_0$ is the characteristic impedance of the line, typically $Z_0 = 50\Omega$. 
Whereas $Z_{tot}$ represents the total impedance, including contributions from the coupling capacitor, the SET, and the impedance transforming circuit:

\begin{equation}
  \label{eq:z_tot}
  Z_{tot} = \frac{1}{j\omega C_C} + j\omega L_C + \frac{R_S}{1+j\omega R_S C_P}.
\end{equation}

Eq. (\ref{eq:z_tot}) describes an ideal impedance transforming network in its simplest form as a lumped element LC-circuit by neglecting the parasitic resistances of the inductor and capacitor $R_L$ and $R_C$. 
This approximation is valid as long as the probing wavelengths exceed the relevant size of the impedance transforming circuit. 
Furthermore, in an oversimplified picture, the SET is modelled as a variable resistor $R_s$. 
The resonance condition for this circuit is given by the condition $Im(Z_{tot} )=0$,

\begin{eqnarray}
  \label{eq:resonant_freq}
  \omega_r = \sqrt{\frac{1}{L_C C_P}(1 - \frac{L_C}{C_P R_S^2})}.
\end{eqnarray}

Eq. (\ref{eq:resonant_freq}) is valid in the limit that the coupling capacitor is much greater than the parasitic capacitance $\frac{C_C}{C_P} \gg 1$.
For typical circuit parameters ($L_C \ll C_P R_s^2$), Eq. (\ref{eq:resonant_freq}) can be further approximated by $\omega_r = \sqrt{\frac{1}{L_C C_P}}$. 
Note that, at resonance, the total impedance is given approximately by, $Z_{tot} (\omega_r )=\frac{L_C}{C_P R_S}$. 
The complex scattering parameters for the conventional reflection setup, $S_{11}$, or the transmission setup, $S_{21}$, can be obtained as \cite{pozar_microwave_2012}:

\begin{eqnarray}
  \label{eq:S21}
  S_{21} = \frac{2}{2+Z_0/Z_{tot}(\omega)},
\end{eqnarray}
\begin{eqnarray}
  \label{eq:S11}
  S_{11} = \frac{Z_{tot} - Z_0}{Z_{tot} + Z_0}.
\end{eqnarray}

\section{\label{sec:level3}Impact of circuit parameters}

This section studies, for both the conventional reflection setup and the above-described transmission setup, the impact of the circuit parameters. 
In particular, the goal is to determine the optimum values of the impedance transforming circuit parameters for a given setpoint of the SET at a high sensitivity point $R_S$ and for a targeted resonance frequency $\omega_r$. 
To investigate the impact of the transforming network circuit parameters we study: 
(1) The signal strength $|S_{11}|^2$ and $|S_{21}|^2$ and
(2) The change in S as a response to changes in $R_S$ by investigating $|\frac{\partial S_{11}}{\partial R_S}|$ and $|\frac{\partial S_{21}}{\partial R_S}|$

\begin{figure}[h]
  \centering
  \setlength\tabcolsep{-4pt}
  \begin{tabular}{c c c}
    \includegraphics[scale=0.4]{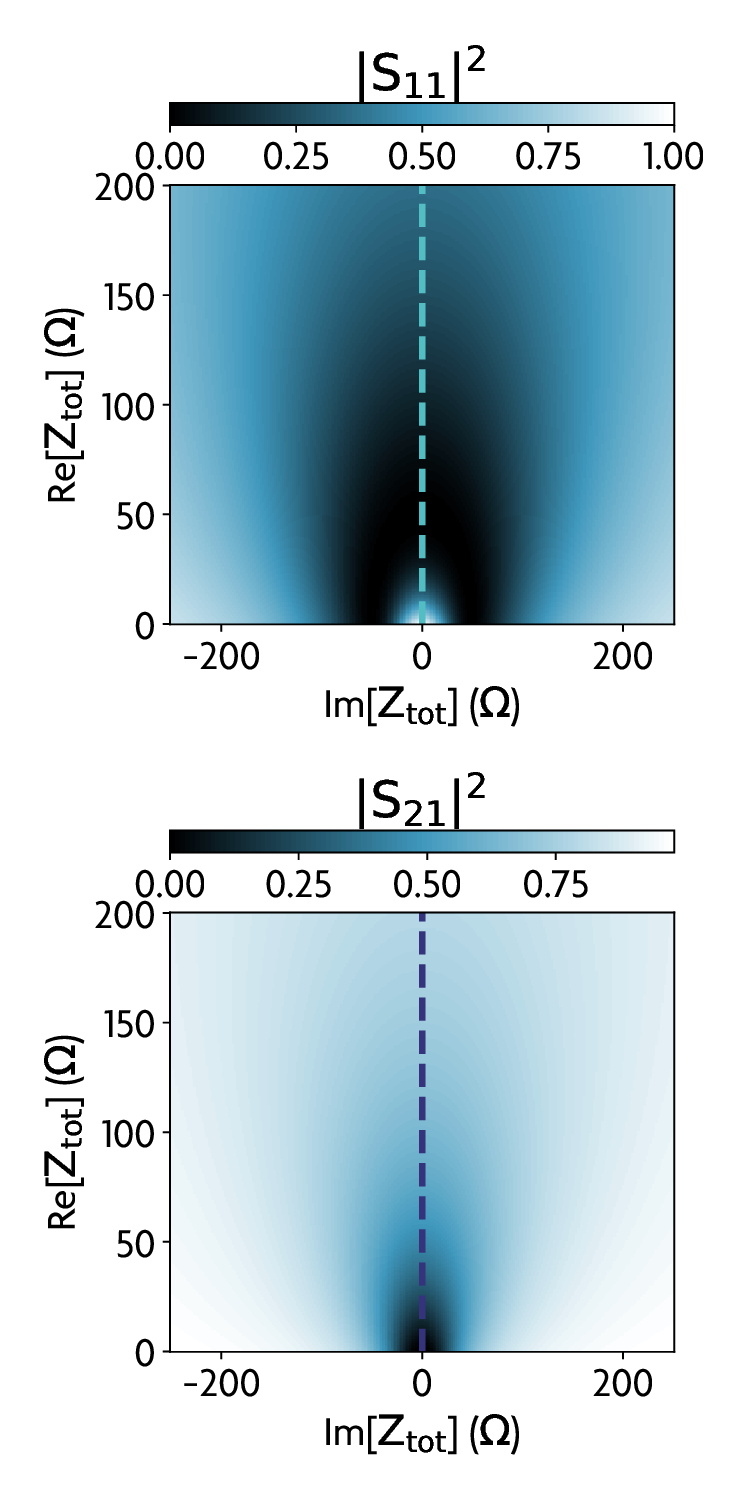}
    &
    \includegraphics[scale=0.4]{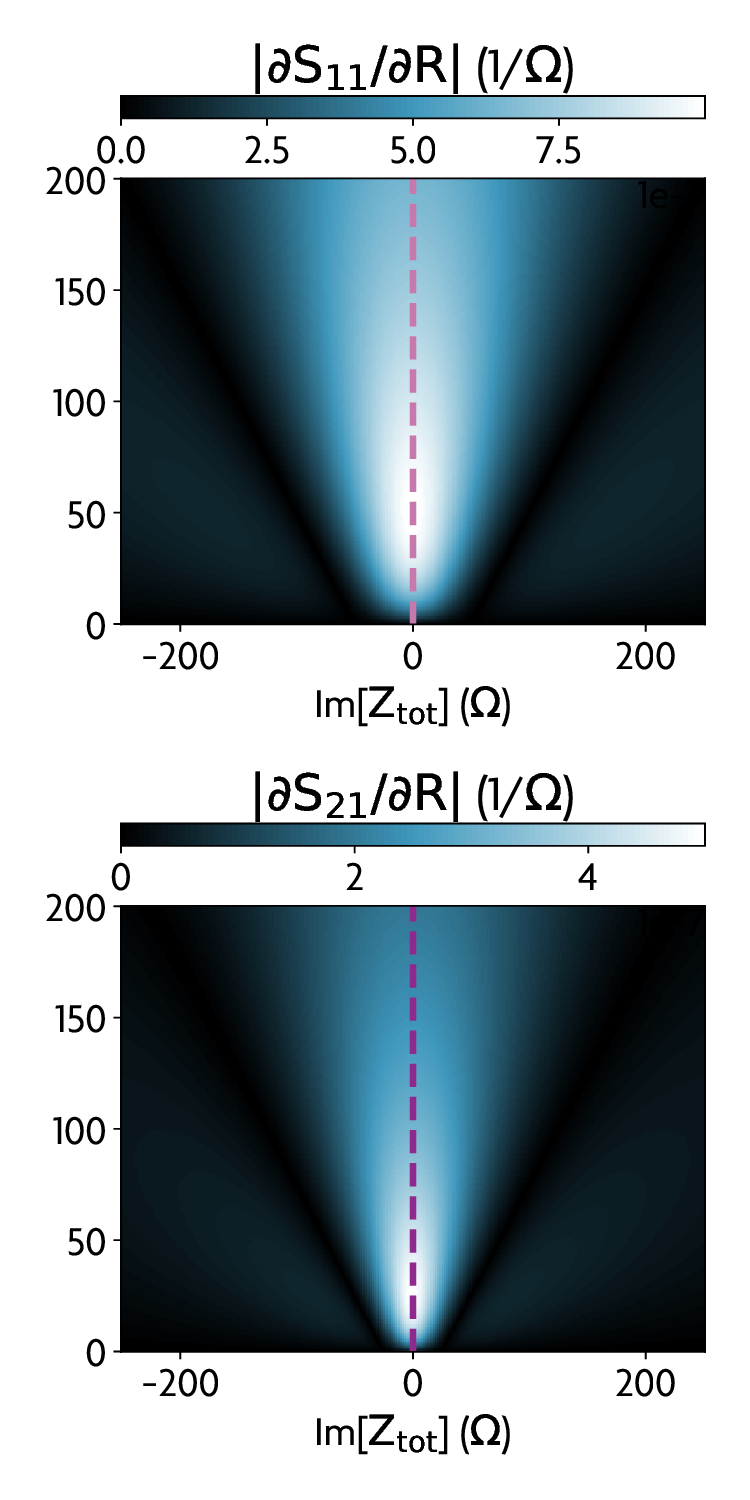}
    &
    \includegraphics[scale=0.4]{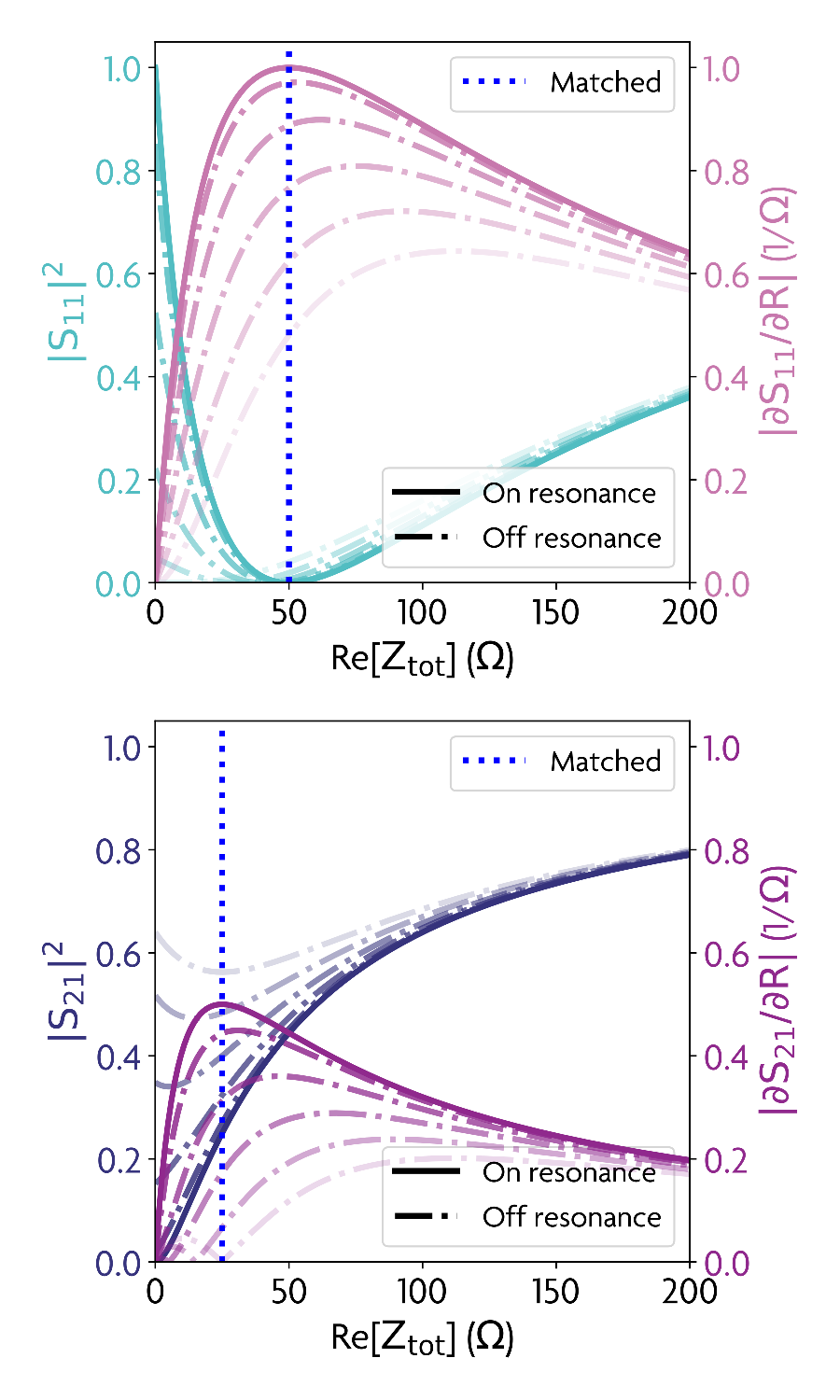}
    \\
    (a) The signal $|S|$ 
    &
    (b) The change in $|S|$

    &
    (c) $|S|$ and $|\frac{\partial S}{\partial R_{S}}|$
    
  \end{tabular}

  \caption{S parameter study of impedance transforming circuit. The top plots are for $S_{11}$, the reflection-based setup and the bottom plots are for $S_{21}$, the transmission-based setup.
  (a) The signal strength with respect to the complex impedance $Z_{tot}$ of the transforming network. The signal is a monotonous function for transmission but for reflection, there is a zero for $|Z_{tot}|=Z_0 (50\Omega)$ 
  (b) The change in $|S|$ as a response to change to $R_S$, $|\frac{\partial S}{\partial R_{S}}|$. The response is maximal along the resonance, $Img(Z_{tot}) = 0$ (dashed lines in a) and b)) 
  (c) Combined plot of $|S|$ and $|\frac{\partial S}{\partial R_{S}}|$ along the resonance (solid lines). $|\frac{\partial S}{\partial R_{S}}|$ is maximal when the circuit is matched to $Z_0(50\Omega)$ in reflection and $\frac{Z_0}{2}(25\Omega)$ in transmission (dotted lines). At this point, $|\frac{\partial S}{\partial R_{S}}|$ is twice as large for reflection than for transmission.}
  \label{fig:rf_set_matching}
\end{figure}

Fig. \ref{fig:rf_set_matching} shows $|S|^2$ and $|\frac{\partial S}{\partial R_S}|$ for both the conventional reflection setup and the above-described transmission setup as obtained from Eqs. (\ref{eq:z_tot} - \ref{eq:S11}), respectively. 
Maximal RF-SET sensitivity is achieved at resonance for $Z_{tot}=Z_0$ when measuring in the conventional reflection setup, and at $Z_{tot}=\frac{Z_0}{2}$ when measuring in a transmission setup (see bottom and top panels in Fig. \ref{fig:rf_set_matching}, respectively). 
At resonance, the maximum response to changes in $R_S$ is twice as large in reflection, and for matched conditions $|\frac{\partial S_{11}}{\partial R_S}| / |\frac{\partial S_{21}}{\partial R_S}| \sim 2$ .

\begin{figure}[h]
  \centering
  \includegraphics[scale=0.5]{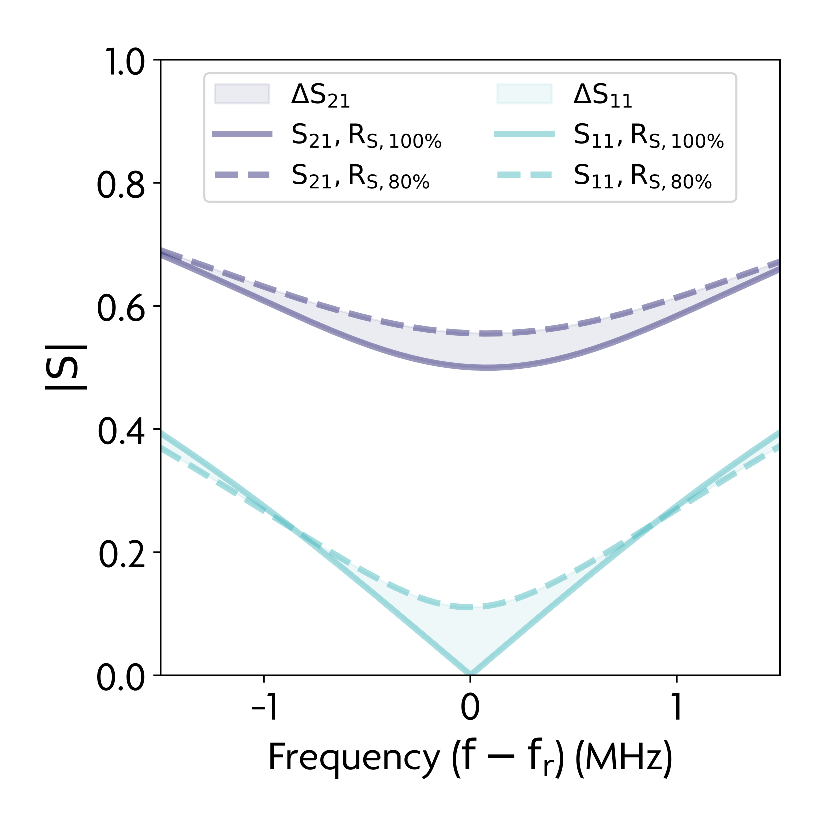}
  \caption{
    Spectrum of the S parameters for reflectometry ($|S_{11}|$, violet) and transmission ($|S_{21}|$, light blue) superimposed along the detuning from resonance. 
    The $|S|$ parameter (solid line) corresponds to a impedance transforming circuit with a coupling capacitance $C_C =100pF$ , SET bias point $R_S =500k\Omega$ and a line impedance of $50\Omega$. 
    $L_C$ and $C_P$ of the impedance transform network are chosen to aim at a common resonance frequency of 350MHz. 
    The $|S|$ parameter change due to a 20\% reduction in $R_S$ are represented as dashed lines. 
    The difference in $|S|$ (lighted area) is twice as large for reflection then for transmission. 
  }
  \label{fig:spectrum}
\end{figure}

To illustrate this further we show in Figure \ref{fig:spectrum} the spectral dependence of $|S_{11}|$ and $|S_{21}|$ versus the detuning from resonance for a 20\% variation (“signal”) of the SET resistance around the SET bias setpoint $R_S$. 
For conventional reflectometry, the circuit parameters, $L_C$ and $C_P$ of the impedance transform network for a given values of the SET bias point $R_S$ ($=500k\Omega$ in Fig. \ref{fig:rf_set_matching}) are calculated with:
\begin{eqnarray}
  \label{eq:cp_reflection}
  C_P = \frac{1}{2\pi f_r \sqrt{R_S Z_0}},
\end{eqnarray}

\begin{eqnarray}
  \label{eq:L_reflection}
  L = Z_0 R_S C_P,
\end{eqnarray}

for a target resonance frequency $f_r$ ($=350MHz$ in Fig. \ref{fig:rf_set_matching}), a coupling capacitance $C_C$ ($=100pF$) and a line impedance $Z_0=50\Omega$.
For transmission, the $L_C$ and $C_P$ are calculated by:

\begin{eqnarray}
  \label{eq:cp_transmission}
  C_P = \frac{1}{2\pi f_r \sqrt{R_S \frac{Z_0}{2}}},
\end{eqnarray}

\begin{eqnarray}
  \label{eq:L_transmission}
  L = \frac{Z_0}{2} R_S C_P.
\end{eqnarray}

From Fig. \ref{fig:spectrum} it is clear that the maximum signal difference is located at resonance and is twice as large for reflection measurements. 
The lower variation of $|S_{21}|$ at resonance for transmission can be partially compensated as transmission allows for a higher input amplitude for the same applied power at device level.

\section{\label{sec:level4}Results}
\subsection{\label{sec:level4:subsec:level1}Multi-module implementation of RF-SET transmission spectroscopy}

Based on the above, we developed a superconductor-semiconductor multi-module microwave assembly to demonstrate radio frequency transmission spectroscopy readout of the charge state of a Si/SiGe semiconductor quantum dot. 
The overall assembly consists of a copper coplanar transmission line located on the measurement PCB, see Fig. \ref{fig:1}(a).

One of the source-drain Ohmic contacts of an SET hosted in a Si/SiGe heterostructure \cite{koch_industrial_2024} is capacitively coupled to the transmission line via a impedance transforming matching circuit and a surface mount bias tee ($C_C=100pF$,$R_B=5.1k\Omega$). 
The PCB allows to couple up to 4 RF-SETs enabling RF multiplexed readout. 
Additionally, the PCB contains $10 k\Omega$ resistors as rf chokes for the gate electrodes. 
The impedance transforming circuit, located in between the coupling capacitor and the Ohmic contact of the SET, consists of a separate 2x2mm high-resistivity silicon die with a 100nm thick Niobium superconducting planar spiral inductor at the surface. 
The spiral inductors are fabricated using a 300mm optical lithography process and a subtractive etching process \cite{verjauw_investigation_2021,van_damme_advanced_2024}. 
As the kinetic inductance of Niobium is low, the designed inductance value can be well approximated by existing analytical expressions for the geometric contribution, with typical errors of 2–3\% \cite{mohan_simple_1999}. 
The superconducting low-loss state of the resonator is beneficial to achieve high internal quality factors. 
The spiral inductors feature two ports, $P_o$ and $P_i$, which allows to connect via wire-bonding the Ohmic contact of the SET (source or drain) to the feedline via $P_o$ and to the coupling capacitor via $P_i$. 
$P_o$ is placed as close as possible to the edge of the chip to reduce the wire-bonding length and parasitic effects as much as possible, while $P_i$ is at the center of the spiral inductor (see Fig. \ref{fig:1}(c)).

The assembly is measured in a Bluefors LD cryostat with a base temperature <10mK. 
The input line to the transmission setup is attenuated by -26dB distributed over the different cryogenic stages, while the output is amplified by 33dB using a cryogenic amplifier (Caltech CITLF3, 10MHz --- 2GHz) mounted at the 4K stage. 
The excitation signal and the demodulated response are generated and measured using a Zurich Instruments UHFLI. 
In a first test, a spiral inductor with an inductance of $L_C=0.5\mu H$ is selected as depicted in Fig. \ref{fig:1}(b).

\subsection{\label{sec:level4:subsec:level2}Impact of micro wave losses on RF-SET transmission reflectometry}

\begin{figure}[h]
  \centering

  \setlength\tabcolsep{0pt}
  \begin{tabular}{c c c}
    \includegraphics[scale=0.4]{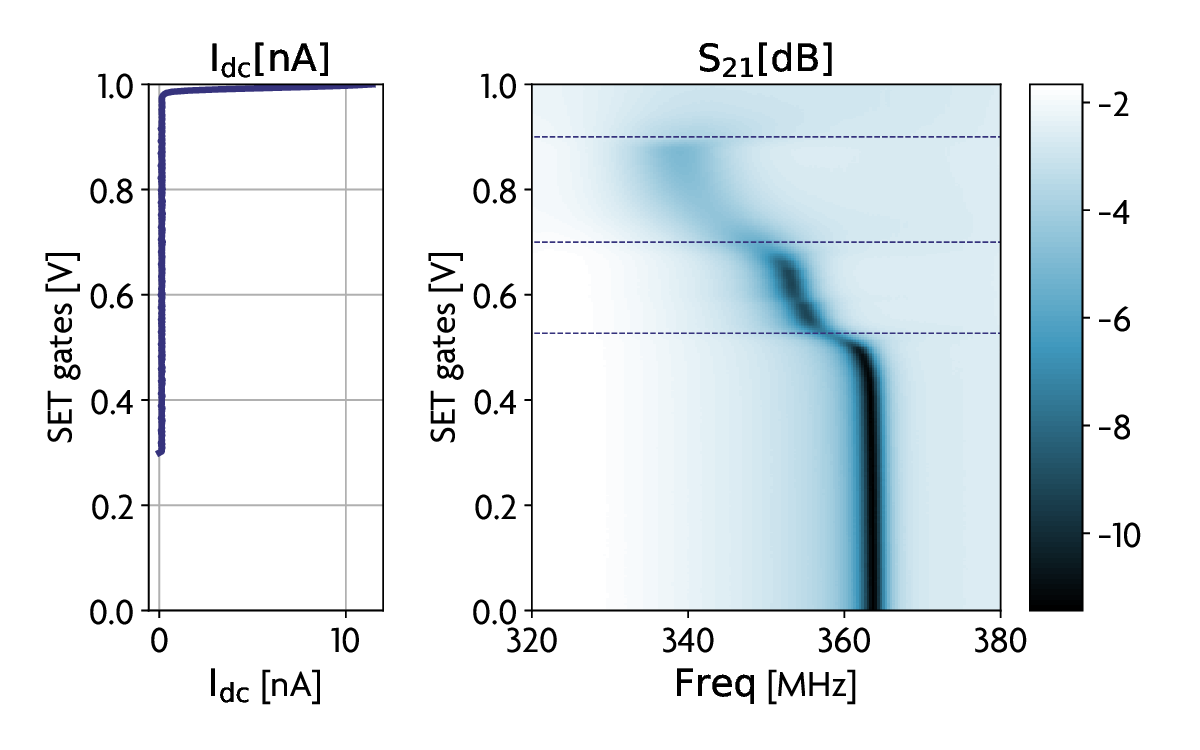}
    &
    \includegraphics[scale=0.34]{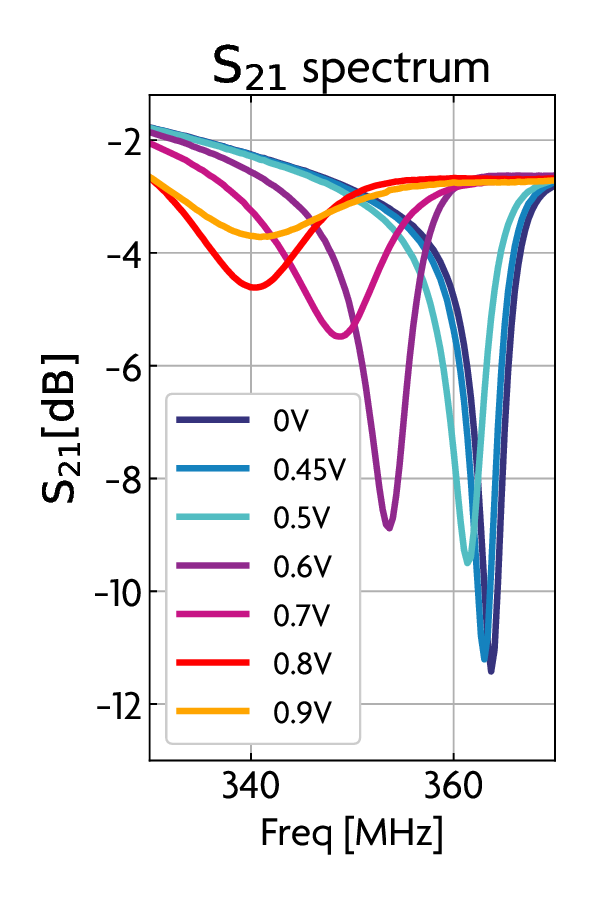}
    &
    \quad\quad\includegraphics[scale=0.042]{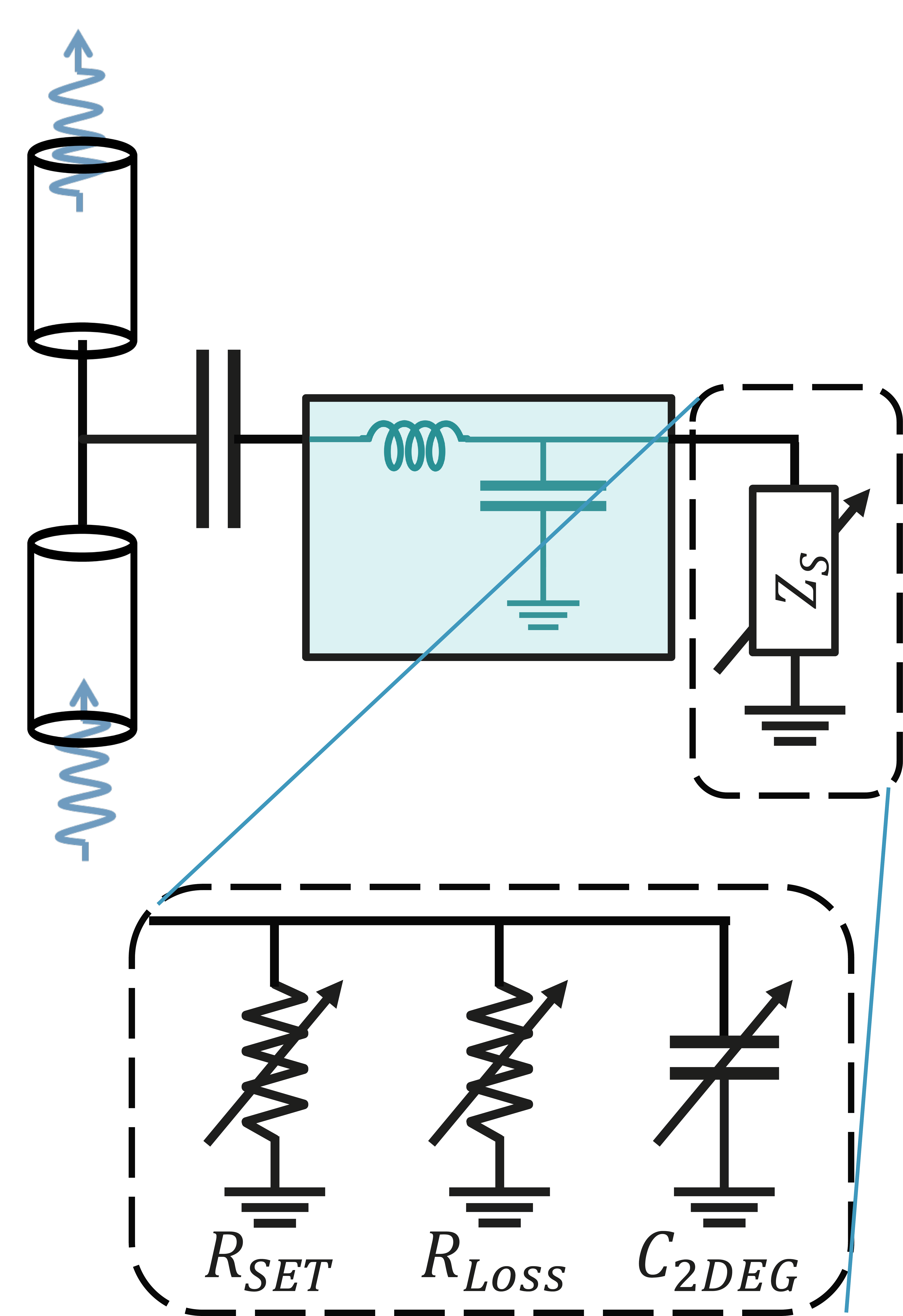}
    \\
    \multicolumn{2}{c}{(a) Global Turn-on}
    &
    (b) Schematic with losses
    \\
  \end{tabular}
  \caption{(a) Global turn-on in RF as well as in DC. There are several distinct regions. 
  The initial part where the 2DEG is not accumulated, with the resonance is located at $f_r= 365MHz$.
  The first frequency shift to $352MHz$ corresponds with an accumulating 2DEG below the AS gate. 
  The second frequency shift to $340MHz$ and an associated strong reduction in quality factor corresponds with further accumulation of a 2DEG below the SET gate. 
  The observed reduction in quality factor in this regime indicates a dissipative path for the RF current. 
  At above 0.9V, the resonance is further suppressed, corresponding with the dc turn-on of the SET.
  (b)A schematic of the transmission setup including an additional gate dependant parasitic capacitance used to quantify the losses via the fitting routine described in the text.
  }
  \label{fig:global_turn_on}
\end{figure}

To start, the RF response to a global turn-on is measured, using a 0.5mVpk RF excitation (all RF values are measured at Lock-in output, unless mentioned otherwise), while dc biasing the SET using a 1mV source-drain bias via the bias tee, see Fig. \ref{fig:global_turn_on}(a). 
A so called global turn-on is performed when all the SET gates (accumulation source/drain (AD,AS), barrier source/drain (BS,BD), plunger (P)) are simultaneously increased, creating a channel through the SET for source-drain transmission.

Several regions can be distinguished. The initial region, with a resonance at around $365MHz$, corresponds to the unaccumulated SET. 
The first frequency shift, at around 0.5V to $353MHz$ matching a capacitive shift of $20.9fF$, can be attributed to an accumulated 2DEG under the AS gate. 
A second frequency, at around 0.7V, to $336MHz$, corresponds to a combined capacitive shift of $59.7fF$ which is of the same order of magnitude as the full accumulated SET's 2DEG capacitance using a parallel plate capacitor model and the design parameters \cite{angus_gate-defined_2007}. 
The observed reduction in quality factor along the turn-on before the perculation threshold is overcome at around 1V cannot be explained with the simple lumped-element model of Section \ref{sec:level3}.
An additional loss mechanism has to be introduced.

Understanding these additional radio frequency losses is important to further develop and optimize RF readout techniques. 
RF losses can result in a strong deterioration of the internal quality factor that cannot be attributed to induced changes in the SET's impedance due to eg. a nearby charge transition.\\ 

These losses can be introduced as a gate dependent parasitic resistor, $R_{Loss}$, in parallel with the device resistance, see Fig. \ref{fig:global_turn_on}(b)\cite{vigneau_probing_2023}. 
The induced capacitance from the accumulated 2DEG, $C_{2DEG}$ can be included in parallel, see Fig. \ref{fig:global_turn_on}(b).
The total capacitance can be rewritten as $C_T$.
In this case, the internal quality factor including the parallel parasitic resistor is given by:

\begin{eqnarray}
  \label{eq:internal_Q}
  Q_i = \sqrt{\frac{C_T R_{eq}^2}{L_C} - 1},\ with\ R_{eq} = \frac{R_{Loss}R_{SET}}{R_{Loss} + R_{SET}} \leq R_{SET}.
\end{eqnarray}

From Eq. (\ref{eq:internal_Q}) it is clear a finite $R_{Loss} \leq R_{SET}$ will have a detrimental effect on the maximum achievable internal quality factor and on the signal swing as a change in $R_{SET}$ will result in a small change of $R_{eq}$ (see Fig. \ref{fig:rf_set_matching}). 
We fit the experimental data with the general equation for the response of a notch port type resonator using the fitting routine as described in Ref. \cite{pozar_microwave_2012}:

\begin{eqnarray}
  \label{eq:s21_Q}
  S_{21} = a e^{j\alpha} e^{-j\omega\tau}(1-\frac{Q_L |Q_C|^{-1}e^{j\phi}}{1+2jQ_L (\frac{f}{f_r}-1)}).
\end{eqnarray}

In Eq. (\ref{eq:s21_Q}), the first factors outside the brackets are due to the environment, whereas the terms inside the brackets are for an ideal resonator. 
$a$ is the initial phase, $\tau$ is the cable delay and $\phi$ is related to the impedance mismatch between the input and output lines. 
$Q_L$ and $Q_C$ correspond to the loaded and coupling quality factor of the circuit. 
From $Q_C$,$Q_L$,$\omega_r$ values extracted from the fit, one can obtain the values of the system $L_C$,$C_T$ and $R_{eq}$.\\

\begin{figure}[h]
  \centering
  
  \setlength\tabcolsep{0pt}
  \begin{tabular}{c c c}
    \includegraphics[scale=0.5]{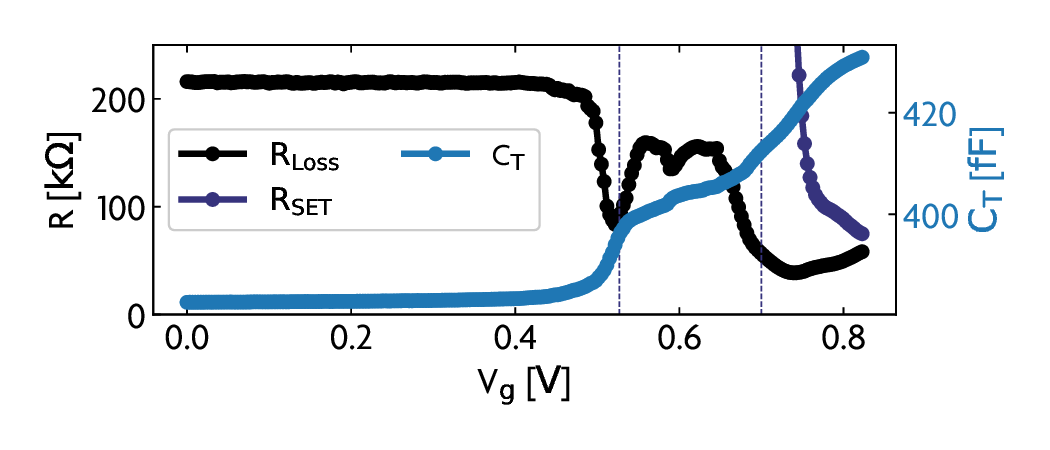}
    &
    \includegraphics[scale=0.3]{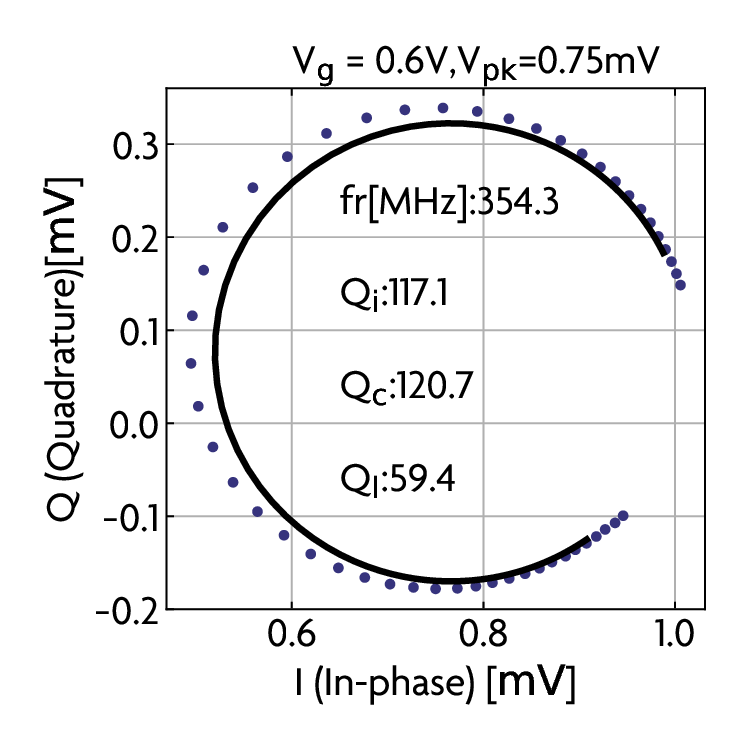}
    &
    \includegraphics[scale=0.3]{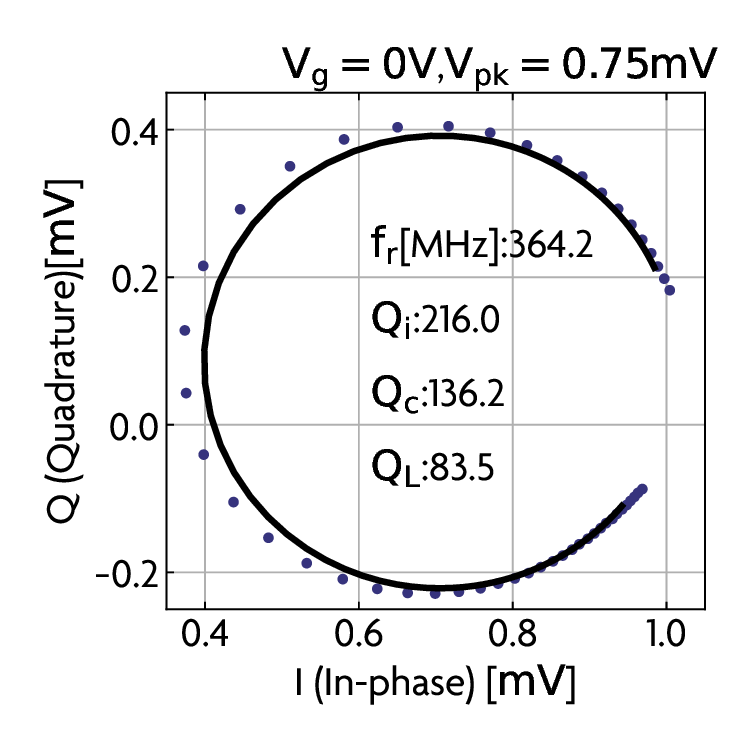}
    \\
    (a) Fitted $R$ and $C_T$
    &
    \multicolumn{2}{c}{(b) Fitted Resonance dips}
      
    \\
  \end{tabular}
  \caption{(a) Evolution of the total capacitive effects ($C_T$), RF losses ($R_{Loss}$) and the SET resistance ($R_{SET}$) along a global turnon. 
  The three regions are visible corresponding to 2DEG accumulations and DC turn-on. When the SET is pinched-off, $R_{Loss}\le R_{SET}$, microwave induced losses are limiting the internal quality factor. 
  When the source gate accumulates, the losses increases as $R_{Loss}$ decreases. Finally, when the percolation threshold is overcome, $R_{SET}$ decreases dramatically with a further reduction in quality factor.
  (b) Resonance dip fitting for two gate voltages, at 0V and 0.6V. The resonance frequencies ($f_r$) and quality factors are extracted.}
  \label{fig:fitting}
\end{figure}

Fig. \ref{fig:fitting}(a) shows the evolution of the different fit parameters during global turn on. The three regions appear again. When the SET is pinched-off, $R_{Loss}\le R_{SET}$, microwave induced losses are limiting the internal quality factor. 
As no 2DEG is accumulated yet, these losses must reside in the vicinity of the Ohmics. 
When the source gate accumulates, the losses increases as $R_{Loss}$ decreases - indicating that an additional loss channel appears due to the accumulated 2DEG. 
Finally, when the percolation threshold is overcome, the SET turns on: there, $R_{SET}$ decreases dramatically, resulting in a further reduction of the quality factor.\\

It is clear that to improve the readout, $R_{Loss}$ must be maximized (losses minimized). 
This can be done by : (i) using high quality low-loss dielectrics and by minimizing the 2DEG surface of the accumulation gates. 
Trap states at the SiGe/Si or SiO$_2$/SiGe interface can interact with the 2DEG, leading to additional scattering and energy loss mechanisms. 
(ii) reducing the losses in the Ohmics. The contact resistance between the Ohmic implants and the 2DEG is a critical source of loss. 
Poorly optimized ohmic contacts can lead to significant power dissipation at high frequencies. 
In Si/SiGe quantum wells, forming low-resistance Ohmic contacts to the 2DEG is challenging due to the need for proper doping and metal-semiconductor interfaces. 
Moreover, Ohmic contacts may suffer from non-idealities, including Schottky barriers or tunnelling resistance that adds to RF losses. 
These unavoidable losses need to be considered when designing the impedance transforming circuit.\\ 

\subsection{\label{sec:level4:subsec:level3}RF-SET transmission spectroscopy for sensing charge transitions in a Si/SiGe quantum dot}

\begin{figure}[h!]
  \centering

  \setlength\tabcolsep{-3pt}
  \begin{tabular}{c c}
    \centering
    \includegraphics[scale=0.4]{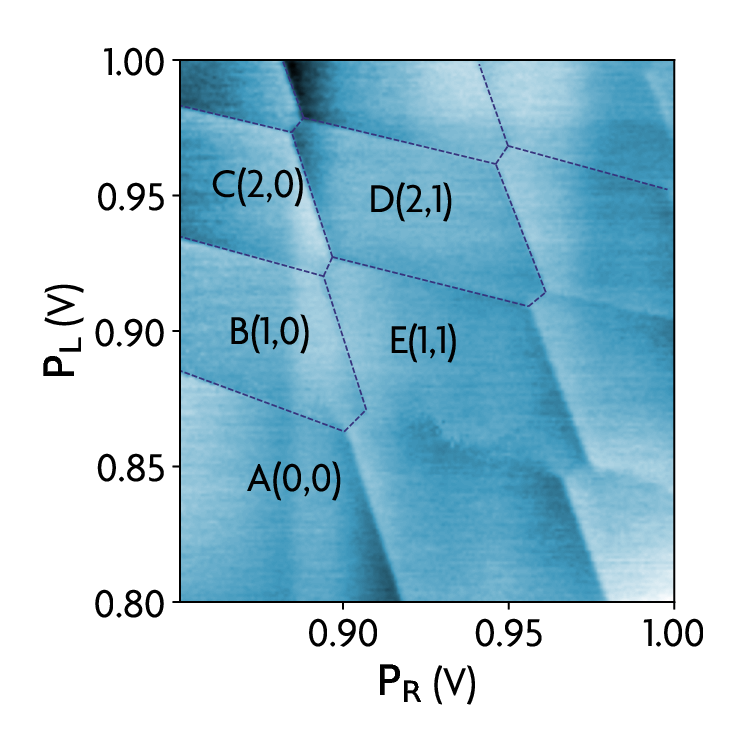}
    &
    \includegraphics[scale=0.4]{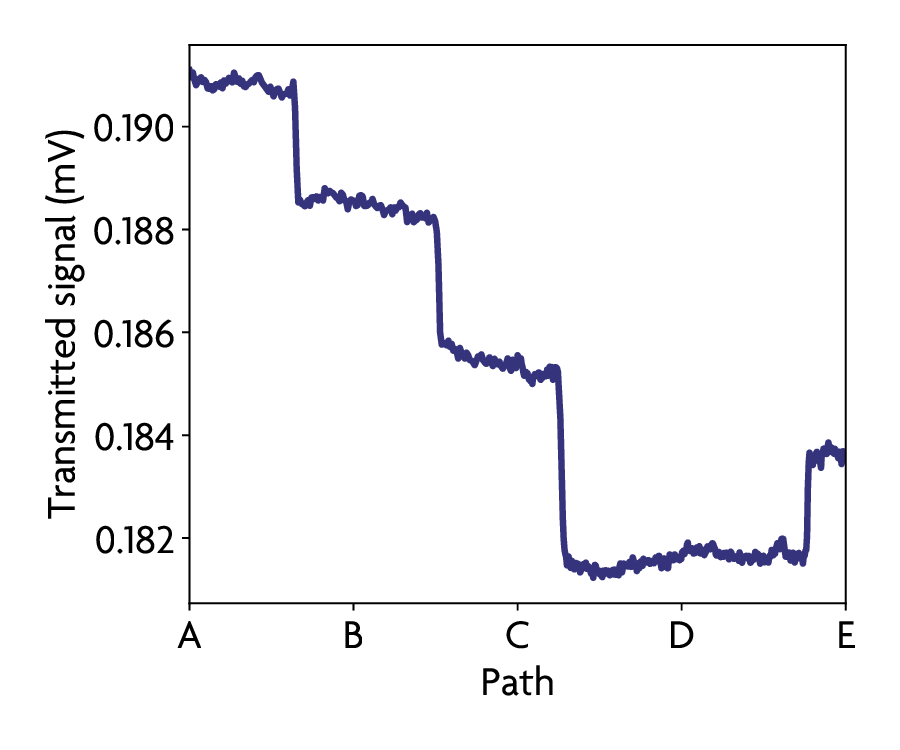}
    \\
    (a) Charge stability diagram
    &
    (b) Compensated path (A to E)
    \\
    \includegraphics[scale=0.4]{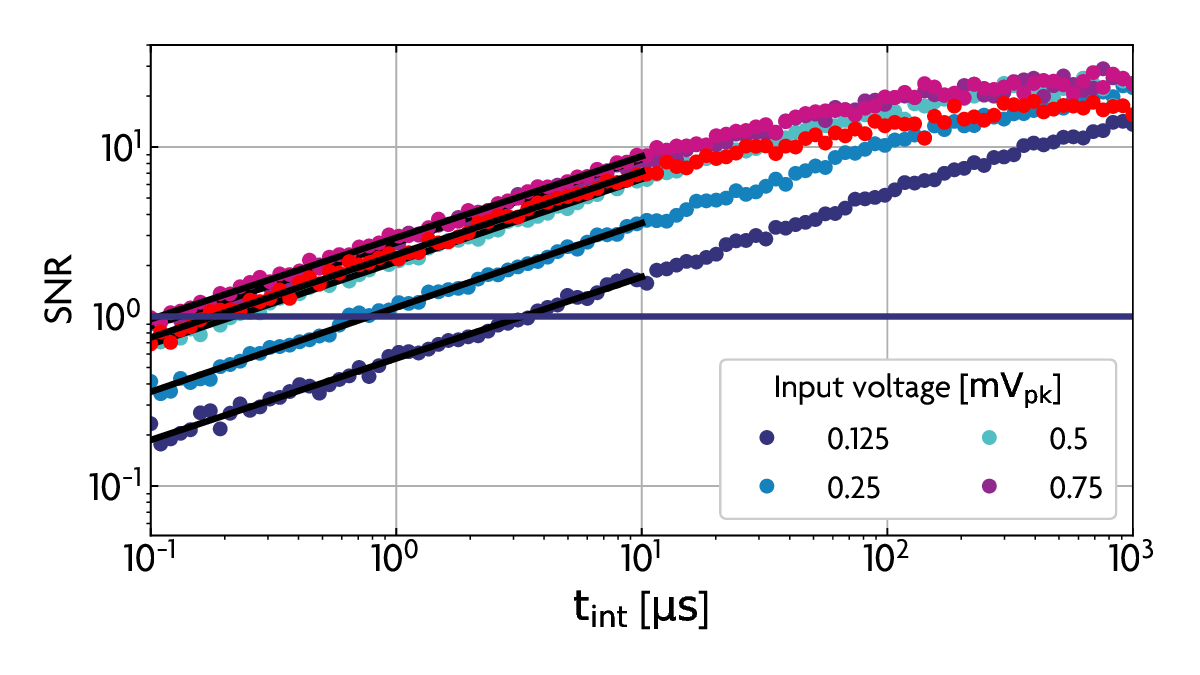}
    &
    \includegraphics[scale=0.4]{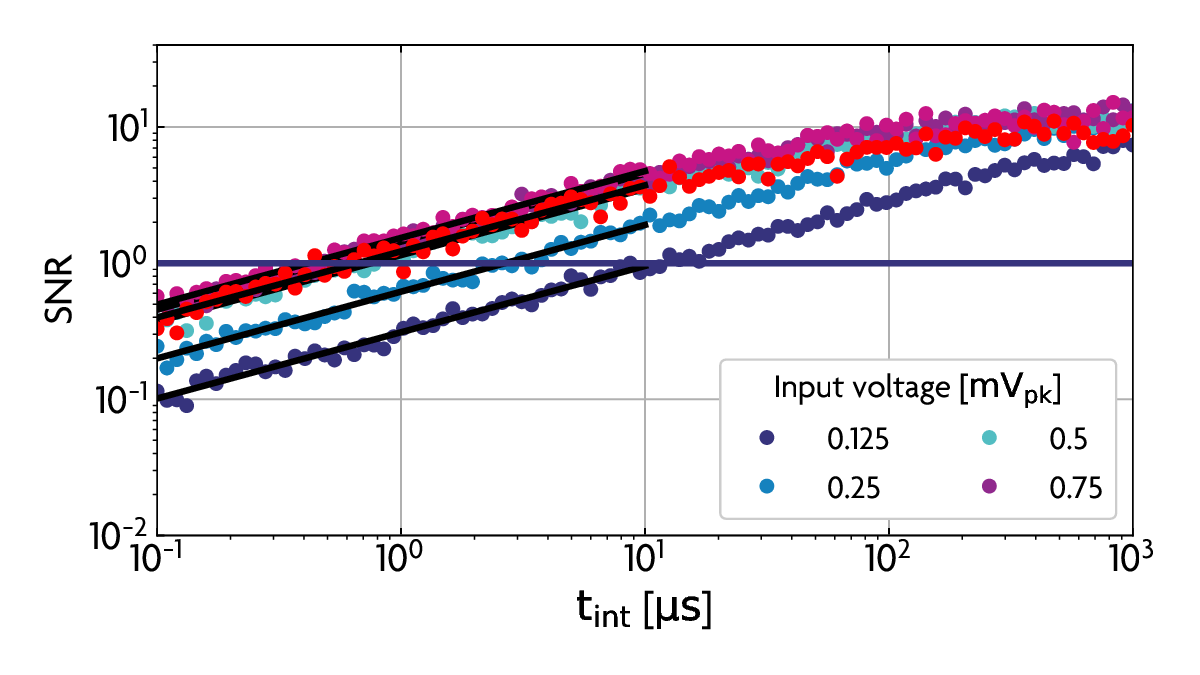}
    \\
    (c) Dot Reservoir transition (DRT)
    &
    (d) Interdot charge transition (ICT)
    \\
  \end{tabular}
  \caption{(a) Measured charge stability diagram of a single quantum dot by monitoring the total response. The map was acquired using an integration time of $t_{int}=1ms$ and a drive amplitude of 0.25mVp. 
  The different charge transitions are indicated by a blue stripe line. 
  (b) The compensated path showing the charge transitions between charge configurations (A to E) 
  (c-d) SNR of a DRT (C to D), resp. ICT (C to E) for different integration times and drive amplitudes. N=2000 points are measured for every power, integration and charge state configuration. 
  The SNR is then calculated between two configurations from a Gaussian fit. The black lines are a fit to $SNR^\alpha  \propto t_{int}$. 
  We reach a SNR=1 for $t_{min}\le1us$ for a suitable power, comparable to current reflectometry set-ups.}
  \label{fig:SNR_study}
\end{figure}

With the above understanding and potential for further improvements, we move to testing the transmission RF-SET on a monolithically integrated double quantum dot device, also based on Si/SiGe technology. 
Fig. \ref{fig:SNR_study}(a) shows a charge stability map of the double dot formed under plungers $P_1$ and $P_2$. 
The charge stability map is measured by tracking the RF response of the SET while it is biased at a high sensitivity point of one of the Coulomb peaks. 
Drifts due to cross-capacitance effects are removed in post-processing of the data.\\

To obtain the SNR versus integration time, $t_{int}$, we acquire N=2000 measurements of the IQ-values on different points on the path A-B-C-D-E (as indicated in Fig. \ref{fig:SNR_study}(a),(b)), crossing different charge configurations. 
We acquire at each point along the path 2000 measurements of the I and Q values for different values of the integration time $t_{int}$. 
The resulting distributions in the IQ plane are fitted using a gaussian distribution. 
The SNR of a transition, for a given integration time $t_{int}$, is then given by,\\

\begin{eqnarray}
  \label{eq;SNR}
  SNR = \frac{|\mu_1 - \mu_2|}{\sqrt{0.5(\sigma_1^2 + \sigma_2^2)}},
\end{eqnarray}

where $\mu_i$,$\sigma_i$ are the mean and average standard deviation of each (i=1,2) of the IQ spots, respectively. 
The SNR is calculated for two transitions, an interdot charge transition (ICT) (C to E) and a dot-reservoir transition (DRT) (C to D), see Fig. \ref{fig:SNR_study}(c),(d). \\

For integration times below $t_{int}= 100\mu s$, the $SNR\propto t_{int}^{1/\alpha}$, with $\alpha\approx2$, corresponds to a dominant contribution of white noise and has a non-monotonic dependence on amplitude\cite{angus_gate-defined_2007}. 
At long integration times, $t_{int}\geq 100\mu s$, the SNR becomes independent of the integration time and drive amplitude corresponding with a dominant 1/f noise as expected at longer time scales. 
For a DRT (ICT) transition, a minimum integration time $t_{min} = 0.1\mu s\ (1\mu s)$ for SNR=1 can be achieved, respectively. 
These values are comparable to state-of-the-art resistive SET readout reported values\cite{vigneau_probing_2023}. 
The difference in $t_{min}$ for ICT and DRT transitions can be directly related to the difference in dipole moment in both transitions. 
Furthermore, we did not observe any impact on the SNR values for magnetic fields of up to 0.5T, applied parallel to the plane of the spiral inductor, demonstrating the field resilience of the resonator.

\section{\label{sec:level5}Conclusion}

In this work, we have demonstrated a superconductor-semiconductor multi-module microwave assembly to demonstrate radio frequency single electron transmission spectroscopy readout. 
We implement the scheme to investigate a Si/SiGe quantum dot device demonstrating a minimum integration time of $t_{min} = 0.1\mu s (1\mu s)$ for a DRT (ICT) transition, respectively. 
The effects of microwave losses on the performance of RF-SET are studied and quantified, which are important for further optimizing the readout. 
Furthermore, parasitic effects due to accumulation style gates have been observed. 
These effects are essential as input for selecting the impedance transforming network parameters to dramatically increase the available SNR\cite{noiri_radio-frequency-detected_2020}, which will be further investigated and optimised in our future work. 
The presented multimodule approach is easily implemented experimentally, shows potential for multiplexed readout of spin qubits\cite{hornibrook_frequency_2014,serrano_improved_2024} and allows studies of rapid charge dynamics in a variety of interesting platforms\cite{connors_charge-noise_2022,keith_single-shot_2019}.

\clearpage
\bibliography{main}

\end{document}